\documentclass[aps, prd, reprint, twocolumn, groupedaddress]{revtex4-1}
\usepackage{amsmath}
\usepackage{amssymb}
\usepackage{graphicx}
\usepackage{color}
\usepackage{times}
\usepackage{euscript}
\usepackage{amsthm}
\usepackage{amsfonts}
\usepackage[english]{babel}
\usepackage{epstopdf}
\usepackage{multirow,makecell,array}
\usepackage{mathtools}
\usepackage{float}

\begin{document}

\title{Photon emission in strong fields beyond\\ the locally-constant field approximation}

\author{I.~A.~Aleksandrov$^{1}$}\author{G.~Plunien$^{2}$} \author{V.~M.~Shabaev$^{1}$}
\affiliation{$^1$~Department of Physics, Saint Petersburg State University, 7/9 Universitetskaya Naberezhnaya, Saint Petersburg 199034, Russia\\$^2$~Institut f\"ur Theoretische Physik, TU Dresden, Mommsenstrasse 13, Dresden D-01062, Germany
\vspace{0.4cm}
}

\begin{abstract}
We investigate a fundamental nonlinear process of vacuum photon emission in the presence of strong electromagnetic fields going beyond the locally-constant field approximation (LCFA), i.e., providing the exact treatment of the spatiotemporal inhomogeneities of the external field. We examine a standing electromagnetic wave formed by high-intensity laser pulses and benchmark the approximate predictions against the results obtained by means of a precise approach evaluating both the tadpole (reducible) and vertex (irreducible) contributions. It is demonstrated that the previously used approximate methods may fail to properly describe the quantitative characteristics of each of the two terms. In the case of the tadpole contribution, the LCFA considerably underestimates the number of photons emitted for sufficiently high frequency of the external field. The vertex term predicts emission of a great number of soft photons whose spectrum is no longer isotropic in contrast to the LCFA results. A notable difference among the photon yields along different spatial directions, which is not captured by the LCFA, represents an important signature for experimental studies of the photon emission process. Since this feature takes place unless the Keldysh parameter is much larger than unity, it can also be used in indirect observation of the Schwinger mechanism.
\end{abstract}

\maketitle

\section{Introduction}\label{sec:intro}

It is well known that Maxwell's Lagrangian for electrodynamics leads to an inherently linear theory, and the corresponding superposition principle does not allow one solution of Maxwell's equations to interact with another; i.e., a combination of two classical light waves does not give rise to any additional radiation. However, {\it quantum} electrodynamics (QED) predicts a phenomenon of photon emission due to quantum fluctuations in the presence of strong external backgrounds~\cite{euler_kockel, heisenberg_euler, weisskopf, schwinger_1951}. This process is closely related to the photon-photon scattering via fermionic loops~\cite{euler_kockel, karplus_pr_1950_1951}, but one assumes here that the initial state contains no photons while the quantized electron-positron field interacts with a classical external background and the quantized part of the electromagnetic field whose quanta are being emitted.

\begin{figure}[b]
\center{\includegraphics[width=0.9\linewidth]{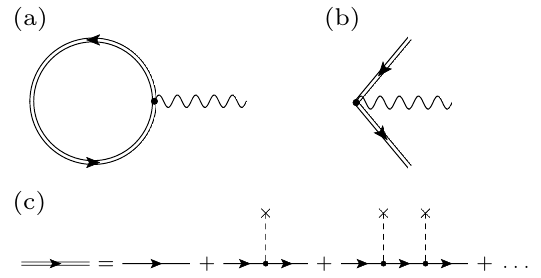}}
\caption{Nonperturbative tadpole (a) and vertex (b) diagrams describing two independent channels of photon emission in a strong external field and giving rise to nontrivial photon number density to first order in the fine-structure constant $\alpha = e^2/(4\pi)$. The double line (c) represents the exact electron propagator (or solution) in the presence of the external field. To calculate the number density of signal photons, one should square the absolute values of the corresponding amplitudes and sum over the final fermionic states in the case of the vertex diagram (b).}
\label{fig:diagrams_gen}
\end{figure}

Although this phenomenon was predicted several decades ago, it has never been investigated experimentally.  Nevertheless, a rapid development of the laser technologies significantly stimulates further attempts at finding most favorable scenarios for practical observations of the effect. From the theoretical viewpoint, it requires new accurate and efficient methods be designed. In particular, the presence of the temporal and spatial inhomogeneities of the external fields in experimental setups demands sophisticated techniques in order to provide adequate predictions for realistic field configurations. In a recent series of studies~\cite{karbstein_prd_2015, gies_prd_2018_1, gies_prd_2018_2, blinne_prd_2019, karbstein_prl_2019}, the authors proposed a very productive computational approach based on the so-called locally-constant field approximation (LCFA) which locally treats the external field as a static and spatially uniform background and invokes the closed-form expression for the Heisenberg-Euler effective action~\cite{heisenberg_euler, schwinger_1951} (it was also utilized in Refs.~\cite{di_piazza_prd_2005, fedotov_pla_2006, fedotov_las_2007, king_pra_2018}). Within this approach, one employs an effective interaction operator defined in the Fock space of photon states and incorporating the one-loop corrections (see, e.g., Ref.~\cite{gies_jhep_2017}). Since this operator does not involve fermionic degrees of freedom, the final expressions turn out to be much less complicated than the exact formulas in terms of the one-particle solutions of the Dirac equation~\cite{fradkin_gitman_shvartsman}. The LCFA approach allows one to efficiently evaluate the tadpole (reducible) contribution [see Fig.~\ref{fig:diagrams_gen}(a)] to the spectra of signal photons taking into account the spatiotemporal dependence of complex field configurations.
However, the validity of the LCFA in context of various processes is always limited. For instance, the LCFA may fail to properly describe the photon spectra in studies of nonlinear Compton scattering~\cite{dipiazza_2018, blackburn_2018, ilderton_pra_2019}. We also point out that according to recent results of precise calculations~\cite{aleksandrov_prd_2017_2, aleksandrov_prd_2018, aleksandrov_prd_2019, kohlfuerst_prd_2018, kohlfuerst_epjp_2018, lv_pra_2018, peng_arxiv_2018, torgrimsson_2018}, the spatiotemporal inhomogeneities may also play a prominent role in the process of electron-positron pair production. In this paper, we perform exact calculations going beyond the LCFA in order to examine its validity.

Furthermore, it turns out that there exists also another contribution besides the tadpole one [see Fig.~\ref{fig:diagrams_gen}(b)], which appears in the same order of perturbation theory (PT) with respect to the radiative interaction (the photon number density${} \sim \alpha$). We will refer to this term as the vertex contribution (alternatively, the irreducible contribution~\cite{gies_jhep_2017}). To our knowledge, it was calculated only in Ref.~\cite{otto_prd_2017} in the case of a spatially uniform time-dependent electric field. According to Ref.~\cite{otto_prd_2017}, the vertex diagram predicts production of a huge amount of soft photons, which also opens up a possibility for studying nonlinear effects of strong-field QED in experiments in the not-so-distant future. Note that the vertex diagram describes the photon-emission process accompanying creation of $e^+ e^-$ pairs, so experimental measurements of the photon spectra could also allow one to indirectly investigate the Schwinger mechanism of pair production. Since real setups involve external backgrounds depending not only on time but also on the spatial coordinates, it is strongly desirable to examine fields having multidimensional inhomogeneities. Moreover, if the external field depends solely on time, the photons associated with this field carry energy but do not transfer momentum, which could misrepresent the spectra of photons emitted.

The paper is organized as follows. In Sec.~\ref{sec:gen} we present a short derivation of the necessary expressions for the tadpole and vertex contributions to the number density of signal photons. In Sec.~\ref{sec:sw} we discuss how these general expressions can be evaluated in the case of a standing electromagnetic wave and what the LCFA predictions are. In Sec.~\ref{sec:results} we present the results of our numerical computations revealing new patterns beyond the LCFA. Finally, in Sec.~\ref{sec:conclusions} we draw a conclusion.

Throughout the article, we use the units $\hbar = c = 1$. The electron charge is $e = -|e|$.

\section{General expressions} \label{sec:gen}

Our calculations are based on the formalism of the Furry-picture quantization of the electron-positron field in the presence of a classical electromagnetic background~\cite{fradkin_gitman_shvartsman}. In order to incorporate photons into this approach, one has to take into account the interaction between the $e^+e^-$ field and the quantized part of the electromagnetic field, whose quanta are being emitted in the process under consideration. We turn to the interaction picture taking into account the interaction operator $H_\text{int} = j_\mu (x) \hat{A}^\mu (x)$ within PT \big[$x = (t, \boldsymbol{x})$\big]. Here $j_\mu (x)$ is the current operator of the electron-positron field, and $\hat{A}^\mu (x)$ is the quantized part of the electromagnetic field. The corresponding $S$ operator reads
\begin{equation}
S = T~\mathrm{exp} \Bigg ( \!\! -i \int \limits_{t_\text{in}}^{t_\text{out}} \! H_\text{int} (t) dt \Bigg ). \label{eq:S_def}
\end{equation}
We assume that the external field vanishes outside the interval $t \in [t_\text{in},\ t_\text{out}]$. The operator~(\ref{eq:S_def}) has the following PT series~\big[$S = S^\text{(0)} + S^\text{(1)} + ... \, {}$\big]:
\begin{eqnarray}
S^\text{(0)} &=& 1, \label{eq:S_0} \\
S^\text{(1)} &=& -i \int \limits_{t_\text{in}}^{t_\text{out}} \! dt \, H_\text{int} (t),
\label{eq:S_1} \\
S^\text{(2)} &=& \frac{(-i)^2}{2} \int \limits_{t_\text{in}}^{t_\text{out}} \! dt_1 \int \limits_{t_\text{in}}^{t_\text{out}} \! dt_2 \, T \big [ H_\text{int} (t_1) H_\text{int} (t_2) \big ],
\label{eq:S_2} \\
&\text{...}& \notag
\end{eqnarray}
The quantized part of the electromagnetic field is decomposed according to
\begin{equation}
\hat{A}_\mu (x) = \sum_{\lambda=0}^3 \int \! d\boldsymbol{k} \, \Big [ c_{\boldsymbol{k}, \lambda} f_{\boldsymbol{k},\lambda, \mu} (x) + c^\dagger_{\boldsymbol{k}, \lambda} f^*_{\boldsymbol{k},\lambda, \mu} (x) \Big ],
\label{eq:photon_field}
\end{equation}
where $c^\dagger_{\boldsymbol{k}, \lambda}$ and $c_{\boldsymbol{k}, \lambda}$ are the photon creation and annihilation operators, respectively, and $f_{\boldsymbol{k},\lambda,\mu} (x) = (2\pi)^{-3/2} (2k^0)^{-1/2} \, \mathrm{e}^{-ikx} \varepsilon_\mu (\boldsymbol{k}, \lambda)$ is the photon wave function corresponding to momentum $\boldsymbol{k}$ ($k^0 = |\boldsymbol{k}|$) and polarization $\lambda$. The electron-positron field operator can be expanded either in terms of the so-called {\it in} one-particle solutions ${}_\pm \varphi_n (x)$ or in terms of the {\it out} solutions ${}^\pm \varphi_n (x)$. The {\it in} ({\it out}) functions are determined by their form at the time instant $t = t_\text{in}$ ($t = t_\text{out}$), where each of them has a certain sign of energy denoted by $\pm$. Quantum number $n$ incorporates momentum and spin. In what follows, we will need the expansion of the electron-positron field operator in terms of the {\it in} solutions of the Dirac equation,
\begin{equation}
\psi (x) = \sum_{n} \big [a_n \, {}_+ \varphi_n (x) + b^\dagger_n \, {}_- \varphi_n (x)\big ], \label{eq:psi_x_in}
\end{equation}
where we have introduced the electron (positron) creation and annihilation operators $a^\dagger_n$ ($b^\dagger_n$) and $a_n$ ($b_n$), respectively. These operators satisfy the usual anticommutation relations. The corresponding vacuum state will be denoted by $|0,\text{in}\rangle$.

\begin{figure*}
\center{\includegraphics[width=0.95\linewidth]{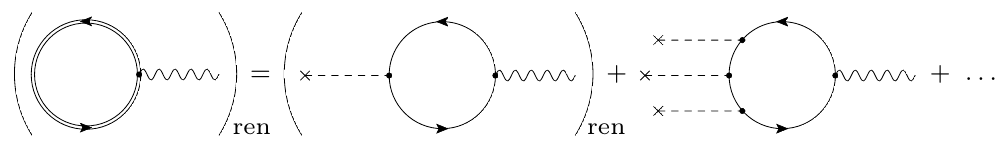}}
\caption{PT expansion of the renormalized tadpole diagram. The dashed lines with crosses denote the interaction with the external field. The renormalized diagram with two external legs does not contribute (see the main text). The diagrams with an odd number of external legs do not appear due to Furry's theorem.}
\label{fig:tadpole_ren}
\end{figure*}

In Ref.~\cite{fradkin_gitman_shvartsman} the authors presented a detailed derivation of transition amplitudes, i.e. $S$-matrix elements, describing the process of photon emission from the vacuum state in the presence of a strong classical external field. Although the corresponding expressions give essentially the leading-order contributions to the photon number density, we will utilize here an alternative approach based on direct calculation of the mean value of the photon number operator (some general aspects and other examples of computing mean values are also discussed in Ref.~\cite{fradkin_gitman_shvartsman}). The exact expression for this quantity \big[$n_{\boldsymbol{k}, \lambda} = dN_{\boldsymbol{k}, \lambda}/d\boldsymbol{k}$\big] reads
\begin{equation}
n_{\boldsymbol{k}, \lambda} = \langle 0,\text{in} | S^\dagger c^\dagger_{\boldsymbol{k}, \lambda} c_{\boldsymbol{k}, \lambda} S |0,\text{in}\rangle.
\label{eq:ph_number_density_int}
\end{equation}
To zeroth order, the field does not generate any photons. Since the interaction Hamiltonian contains terms with only one photon creation/annihilation operator, it is clear that the first-order contribution to the number density~(\ref{eq:ph_number_density_int}) also vanishes. So the main task is to calculate the second-order term given by
\begin{equation}
n_{\boldsymbol{k}, \lambda}^{\text{(2)}} = \langle 0,\text{in} | S^{\text{(1)}\dagger} c^\dagger_{\boldsymbol{k}, \lambda} c_{\boldsymbol{k}, \lambda} S^\text{(1)} |0,\text{in}\rangle.
\label{eq:ph_number_density_second_order}
\end{equation}

In order to perform these calculations, one should express the current operator $j^\mu (x) = (e/2)[\bar{\psi}(x)\gamma^\mu, \, \psi(x)]$ in terms of the {\it in} operators. This can be done by means of Eq.~(\ref{eq:psi_x_in}), so one receives
%
%
\begin{eqnarray}
j^\mu (x) &=& e \sum_{l,s} \big [ ({}_+ \bar{\varphi}_l \gamma^\mu {}_+ {\varphi}_s) a_l^\dagger a_s + ({}_+ \bar{\varphi}_l \gamma^\mu {}_- {\varphi}_s) a_l^\dagger b^\dagger_s \notag \\
{} &-& ({}_- \bar{\varphi}_s \gamma^\mu {}_+ {\varphi}_l) a_l b_s - ({}_- \bar{\varphi}_s \gamma^\mu {}_- {\varphi}_l) b_l^\dagger b_s \big ] \notag \\
{} &+& \frac{e}{2} \sum_{l} \big [ ({}_- \bar{\varphi}_l \gamma^\mu {}_- {\varphi}_l) - ({}_+ \bar{\varphi}_l \gamma^\mu {}_+ {\varphi}_l)\big ],
\label{eq:current_in}
\end{eqnarray}
where the one-particle functions depend on $x$. The term displayed in the third line of Eq.~(\ref{eq:current_in}) represents the {\it in}-vacuum expectation value $j_{\text{in}}^\mu (x) \equiv \langle 0,\text{in} | j^\mu (x) |0,\text{in}\rangle$ (vacuum current). The corresponding contribution will be referred to as the tadpole (reducible) one, while the rest part of $j^\mu (x)$ in Eq.~(\ref{eq:current_in}) will give rise to the vertex (irreducible) term. After some straightforward calculations, one obtains
\begin{equation}
n_{\boldsymbol{k}, \lambda}^{\text{(2)}} = n_{\boldsymbol{k}, \lambda}^{\text{(tadpole)}} + n_{\boldsymbol{k}, \lambda}^{\text{(vertex)}}, \label{eq:ph_number_density_gen_sum}
\end{equation}
where
\begin{widetext}
\begin{eqnarray}
n_{\boldsymbol{k}, \lambda}^{\text{(tadpole)}} &=& \bigg | \int \! d^4x \, j_{\text{in}}^\mu (x) f^*_{\boldsymbol{k},\lambda,\mu} (x) \bigg |^2 = \frac{e^2}{4} \bigg | \sum_{n} \int \! d^4x \, \big [ {}_+ \bar{\varphi}_n (x) \gamma^\mu \, {}_+ \varphi_n (x) - {}_- \bar{\varphi}_n (x) \gamma^\mu \, {}_- \varphi_n (x) \big ] f^*_{\boldsymbol{k},\lambda, \mu} (x) \bigg |^2,\label{eq:gen_tadpole} \\
n_{\boldsymbol{k}, \lambda}^{\text{(vertex)}} &=& e^2 \sum_{n,m} \bigg | \int \! d^4x \, {}_+ \bar{\varphi}_n (x) \gamma^\mu f^*_{\boldsymbol{k},\lambda,\mu} (x) \, {}_- \varphi_m (x) \bigg |^2.\label{eq:gen_vertex}
\end{eqnarray}
\end{widetext}
These are the leading-order contributions which appear in the first order in $\alpha$. They can be represented by means of the Feynman diagrams depicted in Fig.~\ref{fig:diagrams_gen}.

The expression~(\ref{eq:gen_vertex}) for the vertex contribution in the case of a spatially uniform field can also be found in Ref.~\cite{otto_prd_2017}. Although this term can be evaluated directly by means of Eq.~(\ref{eq:gen_vertex}), the tadpole diagram requires renormalization. In Fig.~\ref{fig:tadpole_ren} we display its PT expansion where it is indicated that one should renormalize the term with one interaction vertex, which possesses a quadratical divergence. However, it turns out that it does not contribute if proper renormalization of the electron charge is performed. We note indeed that the amplitude contains the renormalized polarization tensor $\Pi^{\mu \nu} (k) \sim (k^\mu k^\nu - k^2 g^{\mu \nu})$, where $k^2 = 0$ since the signal photon is real. The term with $k^\mu k^\nu$ also vanishes because it is contracted with the photon polarization function, $k^\mu \varepsilon_\mu (\boldsymbol{k}, \lambda) = 0$. Accordingly, the leading contribution is determined by the diagram with four external legs, which can be approximately calculated within the LCFA approach developed in Refs.~\cite{karbstein_prd_2015, gies_prd_2018_1, gies_prd_2018_2, blinne_prd_2019, karbstein_prl_2019}. The leading term is proportional to $(E_0/E_\text{c})^3$, where $E_\text{c}$ is the Schwinger critical value [$E_\text{c} = m^2c^3/(|e|\hbar) = 1.3 \times 10^{16}$~V/cm, where $m$ is the electron mass], while the full tadpole diagram displayed in Fig.~\ref{fig:diagrams_gen}(a) includes also the higher-order terms${} \sim (E_0/E_\text{c})^5$, $(E_0/E_\text{c})^7$, etc. Since the condition $(E_0/E_\text{c})^2 \ll 1$ seems completely realistic, there is no need to evaluate the higher-order terms.

\section{Standing electromagnetic wave} \label{sec:sw}

In the present study, we evaluate both the tadpole and vertex contributions considering a standing electromagnetic wave with peak electric-field strength $E_0$ and frequency $\omega$. The vector potential is chosen in the following form:
\begin{equation}
A_x (t, z) = \frac{E_0}{\omega} \, F(t) \sin \omega t \cos \omega z,
\label{eq:sw_field}
\end{equation}
where $F(t)$ is a smooth envelope function which vanishes unless $t_\text{in} \leqslant t \leqslant t_\text{out}$. The other components of $A^\mu$ equal zero. We always use a large number of cycles ($N \gg 1$). The field configuration chosen has several important advantages. First, this background can be viewed as an approximation for the resulting field of two laser pulses propagating along and opposite the $z$ direction, respectively, and polarized along the $x$ axis \big[the photons carry momentum $\pm \boldsymbol{K}$, where $\boldsymbol{K} = (0, 0, \omega)$\big]. Although this field configuration is infinite in space and does not depend on $x$ and $y$, it is a reasonable approximation for a combination of two counterpropagating laser pulses since $N \gg 1$ and Eq.~(\ref{eq:sw_field}) takes into account the spatiotemporal dependence of the carrier neglecting only slowly varying spatial parts pertaining to the envelope. Second, as $N \gg 1$, the external field frequency related to the temporal oscillations is well defined and coincides with that regarding the spatial dependence in accordance with Maxwell's equations. Finally, the spatial periodicity of the external field allows us to efficiently solve the Dirac equation in the momentum representation, which is necessary for evaluating the number density of signal photons within our approach (see below).


Taking into account the periodicity of the external field~(\ref{eq:sw_field}), one can represent the {\it in} one-particle solutions of the Dirac equation in the following form (see, e.g., Refs.~\cite{bauke_pra_2014, woellert_prd_2015, aleksandrov_prd_2016}):
\begin{equation}
{}_\zeta \varphi_{\boldsymbol{p},s} (x) = (2\pi)^{-3/2} \, \mathrm{e}^{i \zeta \boldsymbol{p} \boldsymbol{x}} \sum_{j=-\infty}^{+\infty} \! {}_\zeta w^j_{\boldsymbol{p},s} (t) \, \mathrm{e}^{i \zeta \omega j z},
\label{eq:sw_solutions_gen}
\end{equation}
where $\zeta = \pm$ and the time-dependent functions ${}_\zeta w^j_{\boldsymbol{p},s} (t)$ are determined by their asymptotic behavior for $t \leqslant t_\text{in}$. One can then demonstrate that the photon number density has the following form
\begin{equation}
\frac{(2\pi)^3}{V} \, n_{\boldsymbol{k}, \lambda}^{\text{(tadpole)}} = \frac{e^2}{(2\pi)^3} \, \sum_l \delta (\boldsymbol{k} - l\boldsymbol{K}) \big | \mathcal{A}_{l, \lambda} \big |^2,\label{eq:tadpole_gen_A}
\end{equation}
where $V$ is the volume of the system and $\mathcal{A}_{l, \lambda}$ are the amplitudes which can, in principle, be calculated to all orders in $E_0 / E_\text{c}$. The delta-function in Eq.~(\ref{eq:tadpole_gen_A}) reflects the momentum conservation law and indicates that the signal photon can be produced after absorbing an integer number of the external-field photons. The energy conservation law does not appear explicitly as we cannot analytically carry out integration over temporal variables. Our calculations revealed that the signal photon is always polarized along the $x$ axis ($\lambda = {\rm x}$). Moreover, the photon yield is substantially suppressed once $l \neq 1$ as the higher-order harmonics appear only in the higher-order terms of PT with respect to $(E_0/E_\text{c})^2$~\cite{fedotov_pla_2006, fedotov_las_2007}. These points are also predicted by the LCFA, which results in the following expression for the leading-order contribution to the photon number density in terms of the function $\mathcal{A}_{l, \lambda}$ for $l=1$ and $\lambda = {\rm x}$:
\begin{eqnarray}
\mathcal{A}^{\text{(LCFA)}}_{l=1, {\rm x}} &=& \frac{\pi}{180} \frac{(e E_0)^3}{m^6} \, m^2 \sqrt{2k^0}  \int \! dt \, \mathrm{e}^{-ik^0 t} \notag \\
{} &\times& \bigg ( 3iQ^3 - Q^2 \, \frac{\dot{Q}}{\omega} - iQ \, \frac{\dot{Q}^2}{\omega^2} + 3 \, \frac{\dot{Q}^3}{\omega^3} \bigg ), \label{eq:tadpole_LCFA_finite}
\end{eqnarray}
where $k^0 = |\boldsymbol{k}| = \omega$ and $Q(t) \equiv F(t) \sin \omega t$. To simplify the computations, we consider an infinite laser pulse \big[$F(t) = 1$, $t_\text{in/out} \to \mp \infty$\big], so the exact result for the leading-order term reads
\begin{equation}
\mathcal{A}^\text{(exact)}_{l=1, {\rm x}} = 2 \pi \delta (k^0 - \omega) \mathcal{I}^{\text{(exact)}},
\label{eq:tadpole_PT_A}
\end{equation}
where
\begin{widetext}
\begin{eqnarray}
\mathcal{I}^{\text{(exact)}} &=& \frac{1}{128} \frac{(e E_0)^3}{m^6} \frac{m^4}{\omega^4} \, m^2 \sqrt{2\omega} \, \sum_{\eta_i, \xi_i} \! \int \! \frac{d\tilde{\omega}}{2 \pi} \int \! d\boldsymbol{p} \, \varepsilon^*_\mu (\boldsymbol{K}, {\rm x}) \notag \\
{} &\times& \mathrm{Tr} \big [ \gamma^\mu S(\tilde{\omega}, \boldsymbol{p}) \gamma^1 S(\tilde{\omega} - \xi_1 \omega, \boldsymbol{p} - \eta_1 \boldsymbol{K}) \gamma^1 S(\tilde{\omega} - (\xi_1 + \xi_2) \omega, \boldsymbol{p} - (\eta_1 + \eta_2) \boldsymbol{K}) \gamma^1 S(\tilde{\omega} - \omega, \boldsymbol{p} - \boldsymbol{K}) \big ],
\label{eq:tadpole_PT_I}
\end{eqnarray}
\end{widetext}
$S(\tilde{\omega}, \boldsymbol{p}) \equiv (\tilde{\omega} \gamma^0 - \boldsymbol{\gamma} \boldsymbol{p} + m)/(m^2 + \boldsymbol{p}^2 - \tilde{\omega}^2)$ is the electron propagator, and the summations run over $\eta_i, \xi_i = \pm 1$ ($i=1$, $2$, $3$) satisfying $\sum \eta_i = \sum \xi_i = 1$. The LCFA prediction~(\ref{eq:tadpole_LCFA_finite}) takes the form
\begin{equation}
\mathcal{I}^{\text{(LCFA)}} = \frac{\pi}{90} \frac{(e E_0)^3}{m^6} \, m^2 \sqrt{2\omega},
\label{eq:tadpole_LCFA_infinite}
\end{equation}
which is to be directly compared with Eq.~(\ref{eq:tadpole_PT_I}). We also point out that as $l=1$ and $\lambda = {\rm x}$, the photons emitted are indistinguishable from those constituting the external laser field (traveling along the $z$ axis). However, our calculations of the tadpole contribution are supposed to examine the accuracy of the LCFA and to survey its justification, which is now possible since Eqs.~(\ref{eq:tadpole_PT_I}) and (\ref{eq:tadpole_LCFA_infinite}) relate to the same physical quantity.

Finally, we note that the leading-order contribution corresponding to the diagram with four external legs (see Fig.~\ref{fig:tadpole_ren}) could contain gauge-dependent (spurious) terms. In atomic physics, where the Coulomb potential of the nucleus is conventionally described by means of the scalar component $A^0$ of the electromagnetic potential, it is a well-elaborated issue. It was demonstrated (see Refs.~\cite{gyulassy, rinker, borie, soff}) that Pauli-Villars regularization~\cite{pauli_villars} uncovers a nontrivial contribution even when the electron mass tends to infinity. This term is gauge-dependent and should be subtracted from the diagram. Nevertheless, in the gauge~(\ref{eq:sw_field}) there is no spurious term (the proof can be found in the Appendix), which was also confirmed by our numerical computations.

As the vertex contribution does not require renormalization, it can be evaluated by means of Eq.~(\ref{eq:gen_vertex}). In the case of a standing wave, plugging the series~(\ref{eq:psi_x_in}) into Eq.~(\ref{eq:gen_vertex}) yields
%
%
\begin{widetext}
\begin{equation}
\frac{(2\pi)^3}{V} \, n_{\boldsymbol{k}, \lambda}^{\text{(vertex)}} = \frac{e^2}{(2\pi)^3} \frac{1}{2k^0} \sum_{s, s'} \int \! d\boldsymbol{p} \bigg | \sum_{j, l} \int \! dt \, {}_+ \bar{w}^j_{-\boldsymbol{p},s} (t) \gamma^\mu \varepsilon^*_\mu (\boldsymbol{k},\lambda) \, {}_- w^{l-j}_{\boldsymbol{p}-\boldsymbol{k}-l\boldsymbol{K}, s'} (t) \mathrm{e}^{ik^0 t} \bigg |^2.
\label{eq:vertex_sw_exact}
\end{equation}
\end{widetext}
When integrating over $t \in (-\infty, t_\text{in}]$ and $t \in [t_\text{out}, +\infty)$, one should introduce a standard factor $\mathrm{e}^{-\varepsilon |t|}$ ($\varepsilon \to 0$) and calculate the integral over these rays analytically (see Ref.~\cite{otto_prd_2017} for more detail). Note that a crucial difference between the tadpole and vertex contributions lies in the conservation laws. Whereas the tadpole term describes the process of photon production by absorption/emission of the external-field photons, the vertex diagram contains also an $e^+ e^-$ pair whose energy can continuously vary. It explains why in the former case the signal photons tend to have the same quantum numbers as the external-background photons (e.g., $k^0 \approx \omega$), while in the latter case they are likely to have a very small energy, $k^0 \ll \omega$, no matter which external-field configuration is chosen. It means first that these two contributions can be analyzed separately. Second, since for a spatially uniform field, Eq.~(\ref{eq:vertex_sw_exact}) becomes much less complicated and the momentum conservation law does not appear to play a vital role here, it seems sensible to perform the calculations replacing the external field~(\ref{eq:sw_field}) with a time-dependent background corresponding to a given position $z$ and to average the results over the spatial period of the standing wave. We will also refer to this approximate technique as the LCFA although the locality is now associated only with spatial coordinates. The predictions of this approach will be benchmarked against those obtained by means of the exact expression~(\ref{eq:vertex_sw_exact}). As was demonstrated in Ref.~\cite{otto_prd_2017}, for small values of $k^0$, the integral over $t \in [t_\text{out}, +\infty)$ in Eq.~(\ref{eq:vertex_sw_exact}) scales as $1/k^0$, which leads to
\begin{equation}
\frac{(2\pi)^3}{V} \, n_{\boldsymbol{k}, \lambda}^{\text{(vertex)}} = \frac{e^2}{(2\pi)^3} \frac{C_\lambda}{(k^0)^3} + \mathcal{O} \bigg (\frac{1}{(k^0)^2} \bigg )
\label{eq:vertex_asym}
\end{equation}
for $k^0 \to 0$. Our calculations proved this asymptotic behavior to remain valid also in the presence of spatial inhomogeneities. In what follows, we will compare the exact results with the LCFA predictions in terms of the coefficient $C = \sum_\lambda C_\lambda$ depending on $E_0$, $\omega$, and the direction of $\boldsymbol{k}$. Since $d\boldsymbol{k} = k_0^2 d \Omega dk_0$, the coefficient $C$ represents the energy density of photons emitted, which turns out to be almost independent of $k_0$, provided $k_0$ is sufficiently small.

To perform exact calculations in the case of a standing wave, we numerically evolve the necessary Fourier components ${}_\zeta w^j_{\boldsymbol{p}, s} (t)$ and evaluate then the expression (\ref{eq:vertex_sw_exact}). Our numerical procedures propagating solutions of the Dirac equation were already successfully employed in several studies concerning $e^+ e^-$ pair production~\cite{aleksandrov_prd_2016, aleksandrov_prd_2017_1, aleksandrov_prd_2017_2, aleksandrov_prd_2018, aleksandrov_prd_2019}. To make sure that we receive reliable data, we first reproduced the results of Otto and K\"ampfer~\cite{otto_prd_2017} and conducted our computations in two different coordinate systems, i.e., we employed Eq.~(\ref{eq:sw_field}) and a similar expression with the substitution $x \leftrightarrow z$.


%
\begin{figure}[t!]
\center{\includegraphics[width=0.95\linewidth]{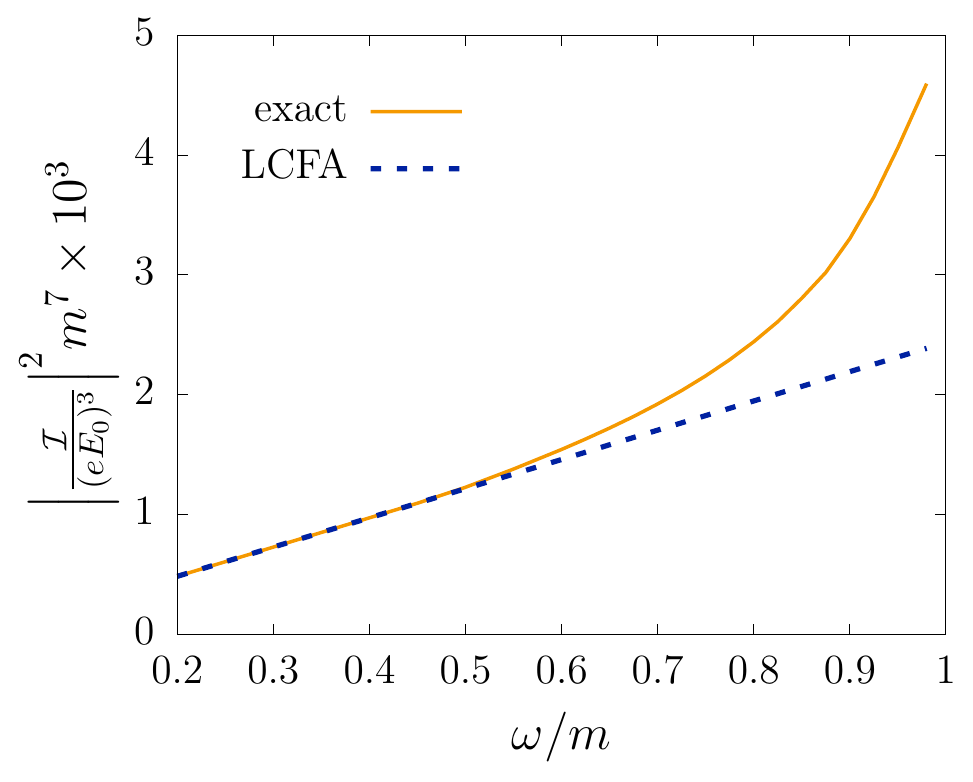}}
\caption{Leading-order contribution to the tadpole diagram evaluated by means of Eq.~(\ref{eq:tadpole_PT_I}) (exact) and Eq.~(\ref{eq:tadpole_LCFA_infinite}) (LCFA) as a function of $\omega$.}
\label{fig:res_four_legs}
\end{figure}

\section{Numerical results} \label{sec:results}

In Fig.~\ref{fig:res_four_legs} we draw a comparison between the exact results and the LCFA predictions for the leading-order contribution to the tadpole diagram in terms of $|\mathcal{I}|^2$ [see Eqs.~(\ref{eq:tadpole_PT_I}) and (\ref{eq:tadpole_LCFA_infinite})] for various values of the carrier frequency $\omega$. The factor $m^6/(eE_0)^3$ is introduced to make the results independent of $E_0$. We observe that for $(\omega / m)^2 \lesssim 0.3$ the LCFA accurately reproduces the exact results, whereas for larger values of $\omega$ it considerably underestimates the photon yield. Although the condition $(\omega / m)^2 \ll 1$ justifying the LCFA was already discussed in the literature (see, e.g., Ref.~\cite{karbstein_prd_2015_2}), the deviation between the LCFA predictions and the exact results was unknown as the latter data has been unavailable until now.

\begin{figure}
\center{\includegraphics[width=0.95\linewidth]{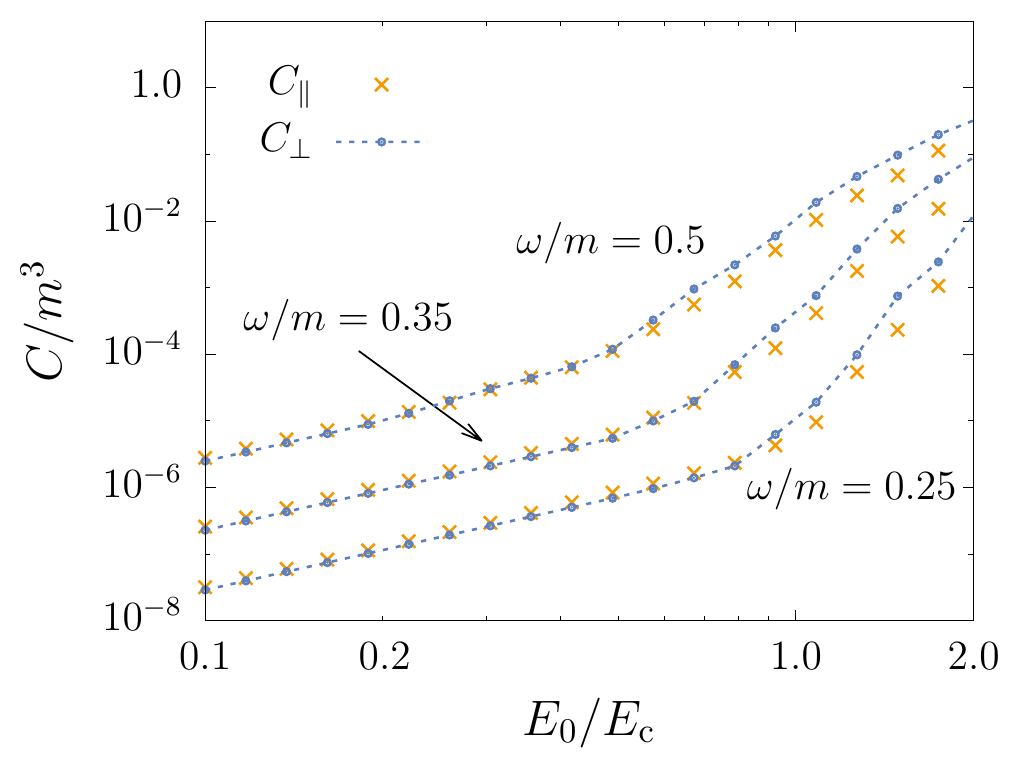}}
\caption{Vertex contribution in terms of the $C$ coefficient evaluated within the LCFA with respect to the parallel direction ($C_\parallel$) and perpendicular one ($C_\perp$) as a function of $E_0$ for various values of $\omega$. ($N=10$).}
\label{fig:res_vertex_lcfa}
\end{figure}

With regard to the vertex contribution, we first underline the fact that the asymptotic behavior~(\ref{eq:vertex_asym}) holds even if one performs the exact computations according to Eq.~(\ref{eq:vertex_sw_exact}). Note also that within the LCFA the $y$ and $z$ axes are completely equivalent, while beyond this approximation all of the three spatial directions are different due to the presence of the magnetic field component. The corresponding $C$ coefficients will be denoted by $C_x$, $C_y$, and $C_z$, e.g., $C_x$ relates to $\boldsymbol{k} = (k^0, 0, 0)$ in Eq.~(\ref{eq:vertex_asym}). Within the LCFA we have $C_x = C_\parallel$ and $C_y = C_z = C_\perp$.

In Fig.~\ref{fig:res_vertex_lcfa} we depict the values of the $C$ coefficients obtained by means of the LCFA. The results reveal two important features. First, one can easily establish the scaling law $C(E_0) \sim E_0^2$ for sufficiently small $E_0$. Moreover, the corresponding threshold value of $E_0$ increases with decreasing $\omega$, so the Keldysh parameter $\gamma = m\omega/|eE_0|$ decreases; e.g., for $\omega = 0.5 m$ it amounts to $\gamma = 1.2$, whereas for $\omega = 0.25m$ it is $\gamma = 0.44$. It means that for $\omega \lesssim 0.1m$, i.e. more realistic frequencies, the scaling law is always valid once the pair-production process, which accompanies photon emission in the vertex diagram, reflects nonperturbative nature, $\gamma \lesssim 1$. This is a very important point because the nonperturbative regime is of particular interest. Second, the function $C(E_0)$ becomes independent of the direction of the photons emitted, i.e. $C_\parallel \approx C_\perp$. Accordingly, the LCFA predicts a great number of soft photons that are emitted isotropically and whose energy density is proportional to $E_0^2$ as long as $\gamma$ is comparable to or larger than unity.

\begin{figure}
\center{\includegraphics[width=0.95\linewidth]{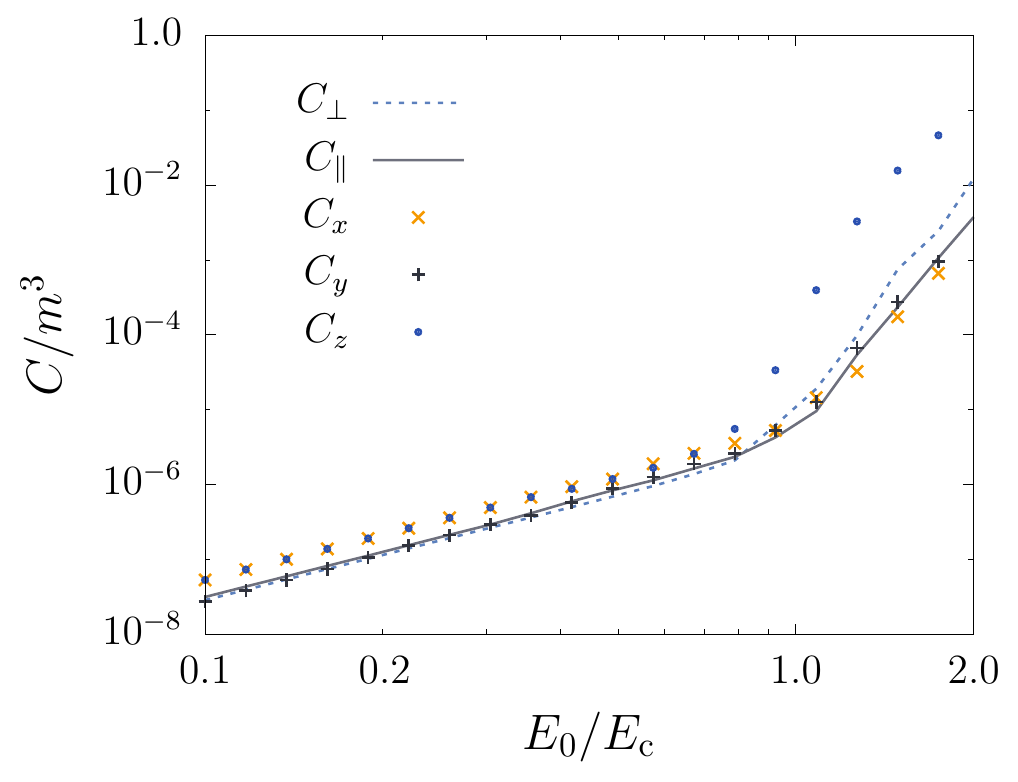}}
\caption{Exact values of the $C$ coefficients corresponding to the three spatial directions $x$, $y$, and $z$ and the LCFA results for $C_\parallel$ and $C_\perp$ as a function of $E_0$ ($\omega / m = 0.25$, $N=10$).}
\label{fig:res_vertex_exact}
\end{figure}

The results obtained beyond the LCFA are displayed in Fig.~\ref{fig:res_vertex_exact} for each of the three spatial directions ($\omega / m = 0.25$, $N=10$). The scaling law $C(E_0) \sim E_0^2$ remains valid, but the emission process is no longer isotropic. Although the curve corresponding to the magnetic field direction $y$ almost coincides with that of $C_\parallel (E_0) \approx C_\perp (E_0)$, for the $x$ and $z$ axes the results are different. While in the former case the coefficient $mC/|eE_0|^2$ amounts to $3\times 10^{-6}$, in the latter case it is two times larger, $6\times 10^{-6}$. These quantitative characteristics predict a notable anisotropy which is expected to be observable in experiment providing a distinctive feature which is not described by the LCFA. The findings discussed above were also confirmed by our computations with other values of $\omega$.

\section{Conclusion} \label{sec:conclusions}

We performed exact calculations of both the tadpole and vertex contributions to the number density of signal photons in the presence of a standing electromagnetic wave. By rigorously treating the temporal and spatial inhomogeneities, we made a first step beyond the previously used approximation. In particular, the results obtained for the tadpole diagram uncovered a substantial underestimation of the photon yield which takes place within the LCFA for sufficiently high frequency of the external field. Concerning the vertex contribution, the exact computations predict a large amount of soft photons whose energy density is proportional to $E_0^2$. More important, the additional radiation is, in fact, anisotropic in contrast to the LCFA results: the number of photons emitted along the $y$ direction is twice as small as the analogous quantity regarding the $x$ and $z$ axes. This fact represents a new important signature which can allow one not only to advance the experimental studies of this phenomenon but also to validate more accurate theoretical approaches going beyond the LCFA. We also emphasize that this finding holds true as long as $\gamma \lesssim 1$, which corresponds to the nonperturbative domain of pair production where detection of photons may open up a possibility of indirect experimental observation of the Schwinger effect. The results of this study are expected to broaden our understanding of nonlinear QED effects supporting further improvement of our knowledge and of the necessary theoretical techniques in pursuit of practical investigations of strong-field QED phenomena.

\begin{acknowledgments}
This work was supported by Russian Foundation for Basic Research (RFBR) and Deutsche Forschungsgemeinschaft (DFG) (Grants No. 17-52-12049 and No. PL 254/10-1) and by Saint Petersburg State University (SPbSU) and DFG (Grants No. 11.65.41.2017 and No. STO 346/5-1). I.A.A. also acknowledges the support from the FAIR-Russia Research Center and from the Foundation for the advancement of theoretical physics and mathematics “BASIS”. V.M.S. also acknowledges the support of SPbSU (COLLAB 2019: No 37722582).
\end{acknowledgments}

\appendix*
\section{Spurious terms of the tadpole diagram} \label{sec:appendix}

As was indicated in the main text, the diagram containing four vertices (see Fig.~\ref{fig:4legs}) could lead to gauge-dependent contributions which should be subtracted. A recipe for extracting these spurious terms is based on the Pauli-Villars regularization procedure~\cite{pauli_villars}. One should replace the electron mass with some mass $M$ and evaluate the diagram assuming $M \to \infty$~\cite{gyulassy, rinker, borie, soff}. The photon wave function corresponding to the external photon line in Fig.~\ref{fig:4legs} is contracted with the current operator $j^{(3)  \mu}_\text{in} (x)$ which has the following form:
\begin{widetext}
\begin{equation}
j^{(3)  \mu}_\text{in} (x) = -ie^4 \! \int \!\! dz_1 \! \int \!\! dz_2 \! \int \!\! dz_3 \, \mathrm{Tr} \, \big [ \gamma^\mu S_M (x, z_1) \gamma^\nu A_\nu (z_1) S_M (z_1, z_2) \gamma^\rho A_\rho (z_2) S_M (z_2, z_3) \gamma^\sigma A_\sigma (z_3) S_M (z_3, x) \big ], \label{eq:current_3_gen}
\end{equation}
\end{widetext}
where the subscript $M$ indicates that the propagators contain now large mass $M$. Since $M \to \infty$, the integrals in (\ref{eq:current_3_gen}) receive nonzero contributions only from the vicinity of $z_1 = z_2 = z_3 = x$, which allows one to replace the arguments of $A$ with $x$. We will also employ the following representation:
\begin{figure}
\center{\includegraphics[width=0.6\linewidth]{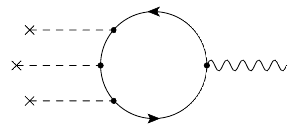}}
\caption{Feynman diagram providing a leading-order term of the tadpole contribution. The solid (fermionic) lines correspond now to the electron propagator involving large mass $M$.}
\label{fig:4legs}
\end{figure}
\begin{equation}
S_M (x ,y) = \int \limits_{C_{\text{F}}} \! \frac{d\omega}{2\pi} \, \mathrm{e}^{-i\omega (x^0 - y^0)} \, g(\boldsymbol{x}, \boldsymbol{y}, \omega),
\label{eq:S_M_representation}
\end{equation}
where $C_{\text{F}}$ denotes the usual contour corresponding to the Feynman propagator,
\begin{equation}
g(\boldsymbol{x}, \boldsymbol{y}, \omega) = \int \!\! \frac{d\boldsymbol{p}}{(2\pi)^3} \, \mathrm{e}^{i \boldsymbol{p} (\boldsymbol{x} - \boldsymbol{y})} \ \frac{\omega \gamma^0 - \boldsymbol{\gamma} \boldsymbol{p} + M}{p_0^2 - \omega^2},
\label{eq:g_definition}
\end{equation}
and $p_0 = \sqrt{M^2 + \boldsymbol{p}^2}$. Using Eqs.~(\ref{eq:S_M_representation}) and (\ref{eq:g_definition}) and integrating over $z_i$ in Eq.~(\ref{eq:current_3_gen}), one obtains
\begin{widetext}
\begin{equation}
j^{(3)  \mu}_\text{in} (x) = -ie^4 A_\nu (x) A_\rho (x) A_\sigma (x) \! \int \! \frac{d\omega}{2\pi} \int \! \frac{d\boldsymbol{p}}{(2\pi)^3} \, \frac{1}{(p_0^2 - \omega^2)^4} \, \mathrm{Tr} \, \big [ \gamma^\mu \Gamma (\boldsymbol{p}, \omega) \gamma^\nu \Gamma (\boldsymbol{p}, \omega) \gamma^\rho \Gamma (\boldsymbol{p}, \omega) \gamma^\sigma \Gamma (\boldsymbol{p}, \omega) \big ], \label{eq:current_3_second}
\end{equation}
\end{widetext}
where $\Gamma (\boldsymbol{p}, \omega) \equiv \omega \gamma^0 - \boldsymbol{\gamma} \boldsymbol{p} + M$. In the case of a nontrivial scalar potential ($A_0 \neq 0$, $\boldsymbol{A} = 0$), this expression yields
\begin{equation}
j^{(3)  0}_\text{in} (x) = - \frac{e^4 A_0^3 (x)}{3 \pi^2} \label{eq:scalar}
\end{equation}
in accordance with Refs.~\cite{gyulassy, rinker, borie, soff}.

If we assume that $A^\mu$ has one nontrivial spatial component, e.g., $A_0 = A_x = A_y = 0$ and $A_z \neq 0$, then the trace in Eq.~(\ref{eq:current_3_second}) gives us odd functions of $p_z$ for each $\mu = 0$, $1$, $2$. Accordingly, integration over $p_z$ yields zero. For $\mu = 3$, one obtains $\mathrm{Tr} \, [ ... ] = 4 ( p_z^4 - 6 a p_z^2 + a^2)$, where $a \equiv M^2 + p_x^2 + p_y^2 - \omega^2$. In this case, the integral over $p_z$ also vanishes:
\begin{equation}
\int \limits_{-\infty}^{+\infty} \!\! dp_z \, \frac{p_z^4 - 6 a p_z^2 + a^2}{(a+p_z^2)^4} = 0. \label{eq:integral_pz}
\end{equation}
Therefore, there is no spurious contribution for the field configuration and gauge chosen.


\end{document}